\documentclass[9pt,twocolumn,twoside]{osajnl}
\journal{ol} % Choose journal (ao, aop, josaa, josab, ol, optica, pr)

\setboolean{shortarticle}{true}
\begin{document}

%%%--------------------------------------------%%%
  % \articletype{Research Article}
  %\received{Month	DD, YYYY}
  %\revised{Month	DD, YYYY}
  %\accepted{Month	DD, YYYY}
  %\journalname{De~Gruyter~Journal}
  %\journalyear{YYYY}
  %\journalvolume{XX}
  %\journalissue{X}
  %\startpage{1}
  %\aop
  %\DOI{10.1515/sample-YYYY-XXXX}
%%%--------------------------------------------%%%

\title{Auto-calibrating Universal Programmable Photonic Circuits: Hardware Error-Correction and Defect Resilience}
%\runningtitle{Integrated Photonic Fractional Convolution Accelerator}
%\subtitle{Insert subtitle if needed}

\author[1,2]{Matthew Markowitz,}
%\ use * to mark the author as the corresponding author
\author[1]{Kevin Zelaya}
%\author[2]{Third Author} 
\author[1,2,*]{Mohammad-Ali Miri}
\affil[1]{\protect\raggedright 
Department of Physics, Queens College of the City University of New York, Queens, New York 11367, USA}
\affil[2]{\protect\raggedright 
Physics Program, The Graduate Center, City University of New York, New York, New York 10016, USA}
\affil[*]{Corresponding author: mmirilab@gmail.com}
%\email{\authormark{*} mmiri@qc.cuny.edu} 
%\communicated{...}
%\dedication{...}

\begin{abstract}
It is recently shown that discrete $N\times N$ linear unitary operators can be represented by interlacing $N+1$ phase shift layers with a fixed intervening operator such as Discrete Fractional Fourier Transform (DFrFT). Here, we show that introducing perturbations to the intervening operations does not compromise the universality of this architecture. Furthermore, we show that this architecture is resilient to defects in the phase shifters as long as no more than one faulty phase shifter is present in each layer. These properties enable post-fabrication auto-calibration of such universal photonic circuits, effectively compensating for fabrication errors and defects in phase components.
\end{abstract}

\maketitle
	
%%%%%%%%%%%%%%%%%%%%%%%%%%  body  %%%%%%%%%%%%%%%%%%%%%%%%%%
\section{Introduction}

Programmable photonic integrated circuits are gaining popularity due to their potential benefits in optical information processing~\cite{harris2018linear, Bogaerts20a, Bogaerts20b}. These platforms have attractive properties such as parallel processing capabilities, lower energy consumption, and higher processing speeds compared to their electronic counterparts. Interest in this topic has been sparked by the realization that arbitrary discrete linear unitary operations can be parameterized into factors that can be represented by simple free-space optical components~\cite{reck1994experimental} as well as analogous integrated photonic circuits \cite{miller2013self}. Since this realization, there has been a flurry of activities on this subject given that it allows for the implementation of arbitrary matrix-vector multiplication on a photonic chip \cite{zhou_photonic_2022, carolan2015universal, ribeiro2016demonstration, taballione20198, Tang2021, taballione2021}. An on-chip photonic matrix-vector multiplier can be broadly deployed in several application scenarios such as in signal processing~\cite{notaros2017programmable}, fiber optic telecommunications, optical neural networks~\cite{shen2017deep}, quantum information and entanglement applications~\cite{harris2017quantum, Wang2019}. Recent studies have shown promising results in this area, and further research is being conducted to advance these technologies.

Fabrication defects and calibration errors can render photonic devices unreliable for immediate applications. Once the device is fabricated, it cannot be modified for error corrections, and external active elements are thus required for calibration. However, there are limits to the amount of calibration that can be done, and sequential calibration is not always possible due to the impact that modifications may have on the device as a whole. In this regard, error mitigation is an essential task when designing photonic architectures. Of particular interest are the architectures capable of optically representing unitary matrices, for those are universal enough to perform arbitrary optical operations. Particular realization of such universal devices are based on meshes of Mach-Zehnder interferometers (MZI) with specific geometries, such as triangular~\cite{reck1994experimental, miller2013self}, rectangular~\cite{clements2016optimal}, diamond ~\cite{Shokraneh2020,rahbardar2023addressing}, as well as hexagonal meshes with protected topological properties~\cite{on2023}. The latter strongly relies on the precision with which MZIs are manufactured, and any defect may render the final device functional. This issue has been recently considered in~\cite{Bandyopadhyay21}, where the authors consider the effects of unitary defect for each MZI. This allows for sequential calibration in meshed architectures, provided certain phases are maintained throughout the calibration process.

Lately, alternative architectures based on cascading of a fixed intervening operator with diagonal phase shift layers, that are capable of representing universal unitary matrices, have been reported in the literature \cite{pastor2021arbitrary, Tanomura20, Tanomura22a, Tanomura23, saygin_robust_2020, Skryabin2021, Markowitz23}. Such an architecture can be realized on-chip with multimode interference couplers \cite{pastor2021arbitrary}, or multicore waveguide couplers \cite{Tanomura22a, Markowitz23}, interlaced with programmable phase shifters. In particular, recently we showed that nonuniform photonic lattices of particularly designed coupling coefficients and length to implement a Discrete Fractional Fourier Transform (DFrFT) operation can be utilized as the intervening operation for realizing programmable unitaries through such an interlacing architecture \cite{Markowitz23}. While a formal proof of the universality of this construction is not currently available, strong numerical evidence, i.e., a phase transition in the norm of representation error, suggests that arbitrary unitary matrices can be realized with remarkable precision, even within the numerical noise error \cite{Markowitz23}. This article focuses on the auto-calibration capabilities of this configuration.

\begin{figure*}[tb]
\centering %\flushleft
\includegraphics[width=0.95\textwidth]{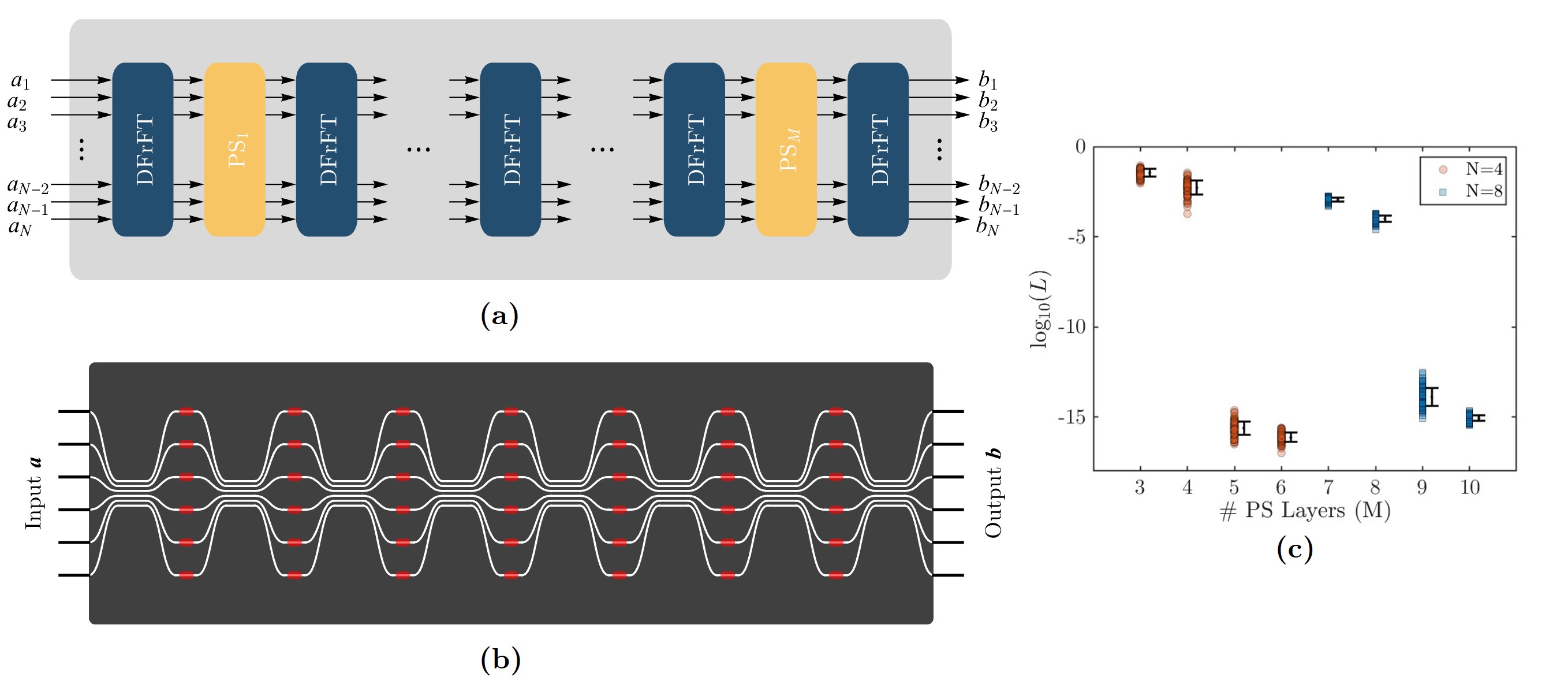}
\caption{(a) The proposed architecture of the N-port system, consisting of alternating layers of Discrete Fractional Fourier Transforms (DFrFT) and diagonal phase shifts layers (PS${}_{j}$). (b) The photonic realization for $N = 6$ using photonic waveguide lattices and phase shifters (red squares). (c) The mean-squared error norms~\eqref{eq_L} of the optimization for $N=4$ and $N=8$, versus the number of phase layers $M$ (for $N=4$, we considered $M=3,4,5,6$ while for $N=8$, we considered 
$M=7,8,9,10$ phase layers).
}
\label{fig1}
\end{figure*}

Figure~\ref{fig1} shows the proposed architecture involving interlacing layers of the Discrete Fractional Fourier Transform (DFrFT) operation and programmable phase shifters. The DFrFT operation can be achieved through the so-called Jx photonic lattice, which is constructed using coupled waveguide arrays with nonunifrom nearest-neighbor coupling rates that ensure equidistant propagation constants of all lattice supermodes~\cite{weimann2016implementation, HonariLatifpour2022}. The universality of such a device has been numerically elucidated~\cite{Markowitz23}. In this work, the auto-calibration properties of the interlaced architecture are analyzed by considering perturbations in the intervening photonic lattices or due to the presence of faulty phase shifters. In the former case, we explore the error induced in the reconstruction of desired unitary matrices due to manufacturing defects in the lattices. Our numerical analysis shows that a second optimization of the phases can bring the error of the reconstructed matrices back to numerical noise levels. In the latter case, we explore reconstruction of target unitary matrices when a number of randomly selected phase shifters are fixed at constant phase values. We find that universality is not jeopardized as long as no two such faulty phase shifters lie in the same layer and the total number of faulty phase shifters is not more than the number of the input/output ports. These characteristics make the interlaced architecture highly adaptable and ensure universality while expanding its potential applications, particularly in large-scale applications where errors must be mitigated as much as possible.

\section{Formulation}
Let us consider a general $N \times N$ unitary transformation matrix $U\in U(N)$, i.e., $U^{\dagger} U = U U^{\dagger} = I$. We are interested in a particular representation for $U$ that allows us to factorize it in terms of other unitary matrices whose optical implementation is feasible. This has been shown to be reliable~\cite{Markowitz23} by considering the interlaced factorization (see Fig.~\ref{fig1}(a))
\begin{equation}
\label{eq_U}
    U = F P_{M} F \cdots P_m \cdots F P_1 F,
\end{equation}
where $P_m$ are diagonal phase matrices with components $P^{(m)}_{p,q}=\delta_{p,q}e^{i\theta_{p}^{(m)}}$, for $p,q=1,\ldots,N$. The superscript $m=1,\cdots,M$ denotes the phase shifter layer index and $\theta_{p}^{(m)}$ the $p$-th phase element in the $m$-th phase shifter layer. In turn, $F$ is the Discrete Fractional Fourier Transform (DFrFT). It is important to note several definitions of the discrete fractional Fourier transform exist in the literature, each based on some on-demand properties imposed \textit{a priori}. See~\cite{Ata97,Candan00,Pei00,weimann2016implementation} for some well-known DFrFT definitions. In this work, we adapt the definition from Ref.~\cite{weimann2016implementation}, for it allows a physical realization of the DFrFT with a particular photonic waveguide array in the form of the so-called Jx lattice. The so-implemented factorization is illustrated in Fig.~\ref{fig1}(b) for $N=6$.

In this fashion, the DFrFT matrix can be written as the propagator generated by such a lattice at the normalized length $\pi/2$. That is,
\begin{equation}
\label{eq_F}
F = e^{i\frac{\pi}{2}H},
\end{equation}
where $H$ is the Jx lattice Hamiltonian~\cite{weimann2016implementation} whose matrix components are $H_{p,q}=\kappa_{p}\delta_{p,q+1}+\kappa_{p-1}\delta_{p,q-1}$, with hopping rates $\kappa_{p} = \frac{\kappa}{2} \sqrt{(N-p)p}$.

Numerical evidence reveals that the interlacing architecture~\eqref{eq_U}, combined with the photonic Jx lattice as the passive matrix $F$, can reconstruct arbitrary $N\times N$ unitary matrices for the appropriate number $M$ of phase shifter layers~\cite{Markowitz23}, i.e., the factorization~\eqref{eq_U} is universal. For completeness, we provide numerical results supporting the universality of~\eqref{eq_U}. We first consider the unperturbed ideal case $F$, and we then optimize the individual phases for an ensemble of randomly chosen target unitary transformations $U_t$, generated in accordance with the Haar measure~\cite{mezzadri_how_2007}. The goodness of approximation of the target matrices is explored against the number of layers $N$, with $M$ phase layers corresponding to $NM$ phase parameters. The loss function is defined as the mean square error
\begin{equation}
\label{eq_L}
L = \frac{1}{N^2} \| U - U_t \|^2,
\end{equation}
where $\|A\|=\sqrt{\text{Tr}(A^{\dagger}A)}$ is the Frobenius norm. We refer to Eq.~(\ref{eq_L}) as the error norm. The optimization is done using the Levenberg-Marquardt algorithm (LMA), which is well suited to sum-of-squares objective functions and which can be used for both under and over-determined problems \cite{levenberg1944method,Marquardt63}. The function tolerance and the step tolerance were set to $10^{-6}$ and the optimality tolerance to $10^{-10}$. Furthermore, 100 target unitary matrices were generated at each matrix size of $N=4,6$ and $N=8$. For a given target, the phases were randomly initialized between $0$ and $2\pi$, and the LMA was run 100 times to find the parameters corresponding to the lowest error norm. Performance tests under these conditions are shown in Figure~\ref{fig1}(c), where the loss function reveals a phase transition from the step tolerance to the optimality tolerance when transiting to $M\geq N+1$ phase shifter layers~\cite{Markowitz23}. Thus, universality is reached for
\begin{equation}
\label{eq_M}
M=N+1,
\end{equation}
phase shifter layers, leading to an over-estimated problem that involves the estimation of $N(N+1)$ free phase parameters for each target matrix. The convergence trend for $M=N+1$ is also supported by the recent numerical results presented in~\cite{Taguchi23} for other layered architectures.

\begin{figure}[h!]
\centering
\includegraphics[width=0.5\textwidth]{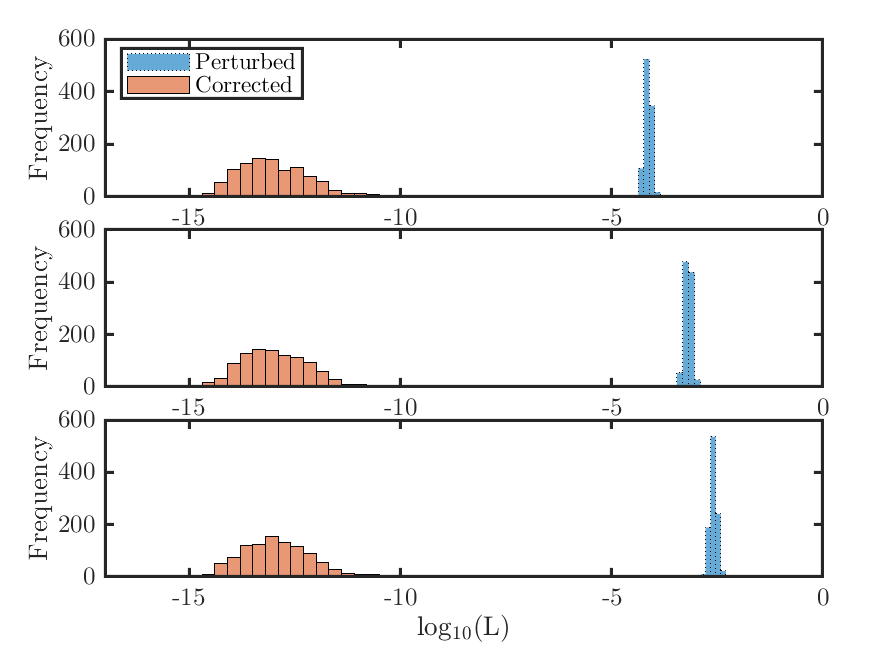}
\caption{The auto-calibration property of the proposed architecture. By perturbing the DFrFT matrices, the error norm jumps to large values, but after a second optimization, new phases are found so that error norms reduce to numerical noise levels. This analysis is done for $(N = 8, M=9)$, and by considering $100$ random target matrices. Here, the perturbation magnitude parameter $\sigma_k$ has been chosen such that the relative error $\Delta F$ is $0.76\%$ (upper row), $2.28\%$ (middle row), and $4.55\%$ (lower row).}
\label{fig2}
\end{figure}

\section{Self-calibration}
In a realistic scenario, the DFrFT matrix $F$ might include some perturbations, e.g., because of fabrication imperfections. In such a case, the factorization~\eqref{eq_U} may render to reconstructed target matrices with a significant error. This cannot be amended once the Jx lattice has been manufactured, and we thus require an alternative mechanism to compensate for any manufacturing errors by external means, such as the phase shifter layers. Although there are various ways to incorporate errors into the interlacing architecture~\eqref{eq_U}, we are only focusing on cases where the construction remains unitary. The most straightforward approach to achieve this is by defining the perturbed lattice Hamiltonian $H_{p}=H+\sigma H_{1}$, where $H_1$ is a Hermitian perturbation matrix with the real and imaginary parts of each entry drawn independently from $\mathcal{N}(0,1)$. The perturbation parameter is taken to be $\sigma = \sigma_k \kappa_{\text{max}}$, where $\kappa_{\text{max}}=\text{max}\{\kappa_{p,p+1}\}_{p=1}^{N-1}$ represents the largest of the coupling coefficients, and $\sigma_{k}$ is a small number so that $\sigma_{k}\kappa_{\text{max}}\ll 1$. The perturbed unitary DFrFT can be constructed as the propagator $F_{p}=e^{i\frac{\pi}{2}H_{p}}$ to ensure its unitary nature. The mean perturbation $\Delta F=\Vert F-F_{p}\Vert/\Vert F \Vert$ has been computed to understand the error caused by the perturbation. The results are shown in the middle column of Table~\ref{table:tab1} for different values of the perturbation parameter $\sigma_k$. It is important to note that even slight values of $\sigma_{k}$ around $10^{-4}$ lead to significant differences in the perturbed matrix $F_{p}$. 

\begin{table}[h!]
\centering % center-align tables within a column
\begin{tabular}[h]{ p{2cm} p{2cm} p{2cm} p{2cm}|  }
 $\sigma_k$ & \% Error F & \% Error U \\
 \hline
 0.001   & 0.76\%    & 2.41\%\\
 0.003 &   2.28\%  & 7.20\%   \\
 0.006 & 4.55\% & 14.40\%\\
\end{tabular}
\caption{The mean perturbation errors in $F$ and the corresponding mean perturbation errors in $U$ (when using uncorrected phases) for different values of $\sigma_k$. Here the mean perturbations are $\Delta F = \|F-F_{p}\| / \|F\|$ and $\Delta U = \|U_t-U_{p}\|/\|U_t\|$, where $F_{p}$ is any one of the $N+1$ perturbed DFrFT matrices, and $U_{p}$ is the transformation matrix using the perturbed DFrFT matrices and the uncorrected phase parameters.}
\label{table:tab1}
\end{table}

Thus, perturbations in $F$ are expected to increase the error accordingly in the reconstruction of the target matrix if proper corrections are not considered. To illustrate the magnitude of this error, we first determine the phase parameters $\theta_{p}^{(m)}$ for a given target matrix $U_t$ and the ideal unperturbed factorization~\eqref{eq_U} using the exact DFrFT matrix $F$ and the LMA optimization scheme. Then, we construct the perturbed target matrix $U_{p}$ through~\eqref{eq_U} by considering the perturbed matrix $F_p$ and the previously computed uncorrected phase parameters $\theta_{p}^{(m)}$. This allows estimating the mean error in the reconstruction process by computing the relative error $\Delta U=\Vert U_{t}-U_{p}\Vert/\Vert U_{t}\Vert$. Numerical results are presented in the right column of Table~\ref{table:tab1}, which indicates that errors induced into $U_{t}$ are about one order of magnitude larger than the errors in $F$. This is expected, for the error in each layer $F$ accumulates throughout the whole factorization~\eqref{eq_U}. 

\begin{figure*}[h!] % h!
\centering %\flushleft
\includegraphics[width=0.92\textwidth]{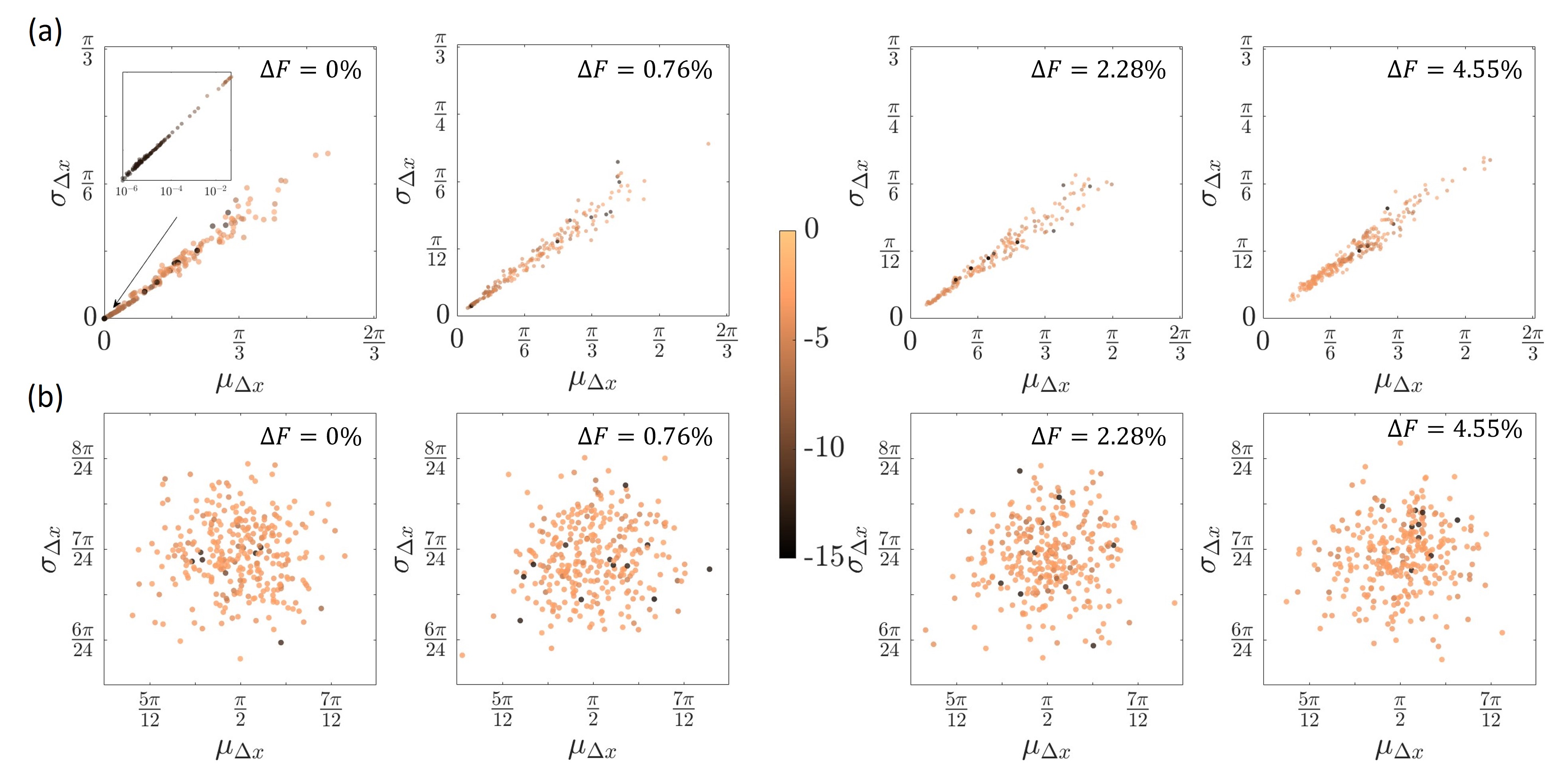}
\caption{Means $\mu_{\Delta x}$ and standard deviations $\sigma_{\Delta x}$ of the difference vector of the original and re-calibrated phases. Each point represents a single run of truncated LMA with 50 iterations for a single perturbed structure and target, with the color representing the norm log$_{10}(L)$. (a) Initial vector chosen within 10\% of the unperturbed vector. (b) Initial vector randomly chosen.}
\label{fig3}
\end{figure*}

In order to account for errors caused by perturbations on $F$, a second optimization process is performed to reconstruct the target matrices $U_{t}$ while considering the perturbed matrix $F_{p}$ in the factorization~\eqref{eq_U}. This optimization leads to a new set of corrected phase parameters $\{\widetilde{\theta}_{p}^{(m)}\}_{p=1,m=1}^{N,N+1}$. For comparison, we compute the loss function~\eqref{eq_L} for randomly generated targets $U_{t}$ and the corresponding reconstructed $U_{p}$ (phase uncorrected) and $\widetilde{U}_{p}$ (phase corrected) target matrices. These results are illustrated in Fig.~\ref{fig2} for $100$ random target matrices, with $N=8$ ports and $M=9$ phase layers. The error obtained when the uncorrected phases $\theta_{p}^{(m)}$ are used is above the established tolerance error $10^{-10}$, whereas the corrected phases $\widetilde{\theta}_{p}^{(m)}$ render for reconstructed matrices with errors below the noise level. In the latter, for each matrix, we search for the optimal phase values $10$ times with different sets of perturbed DFrFT matrices each time. The truncated LMA was used with a maximum of $50$ iterations per run. For the values of the perturbation parameter $\sigma_k$ tested, the truncated LMA was always able to find error norms below $10^{-10}$. This clearly demonstrates that our results on universality do not depend critically on the precise form of $F$. Accordingly, we expect that in a physical realization, fabrication errors in the DFrFT layers can be readily balanced post-fabrication by tuning the reconfigurable phase shifters to achieve a precise realization of a desired unitary matrix.

The error correction during the second optimization process is better illustrated by conveniently introducing the original (uncorrected) and re-calibrated phase vectors $\mathbf{x}=(\theta_{1}^{(1)},\ldots,\theta_{N}^{(1)},\ldots,\theta_{1}^{(N+1)},\ldots\theta_{N}^{(N+1)})$ and $\bar{\mathbf{x}}=(\widetilde{\theta}_{1}^{(1)},\ldots,\widetilde{\theta}_{N}^{(1)},\ldots\widetilde{\theta}_{1}^{(N+1)},\ldots,\widetilde{\theta}_{N}^{(N+1)})$, respectively. One can thus compute the mean $\mu_{\Delta x}$ and standard deviations $\sigma_{\Delta x}$ of the difference vector $\Delta \mathbf{x}=\mathbf{x}-\widetilde{\mathbf{x}}$, such that these quantities capture any deviation during the re-calibration process. Figure~\ref{fig3} shows how the optimized phase vector differs from a given phase vector used to construct a single target matrix. The target matrix is constructed using the unperturbed DFrFT, and the optimization is done both for perturbed and unperturbed DFrFT. In either case, many low-norm solutions can be found. When LMA is run without constraining the initial vector, the re-calibrated phase vectors will statistically be uncorrelated with the given vectors. When each element of the initial vector in the optimization is chosen to be within $\pm 10\%$ of the given vector (but without constraining the parameter space), one can find solutions which are close to the given vector, both when running optimization for perturbed and unperturbed DFrFT. The perturbation parameter $\sigma_{k}$ was appropriately chosen in order to achieve the desired relative errors $\Delta F=\Vert F-F_{p}\Vert/\Vert F \Vert \times 100\%= 0\%,0.76\%, 2.28\%, 4.55\%$ for the tests presented in Figure~\ref{fig3}.

\begin{figure*}[h!] %h!
    \centering
    \includegraphics[width=0.85\linewidth]{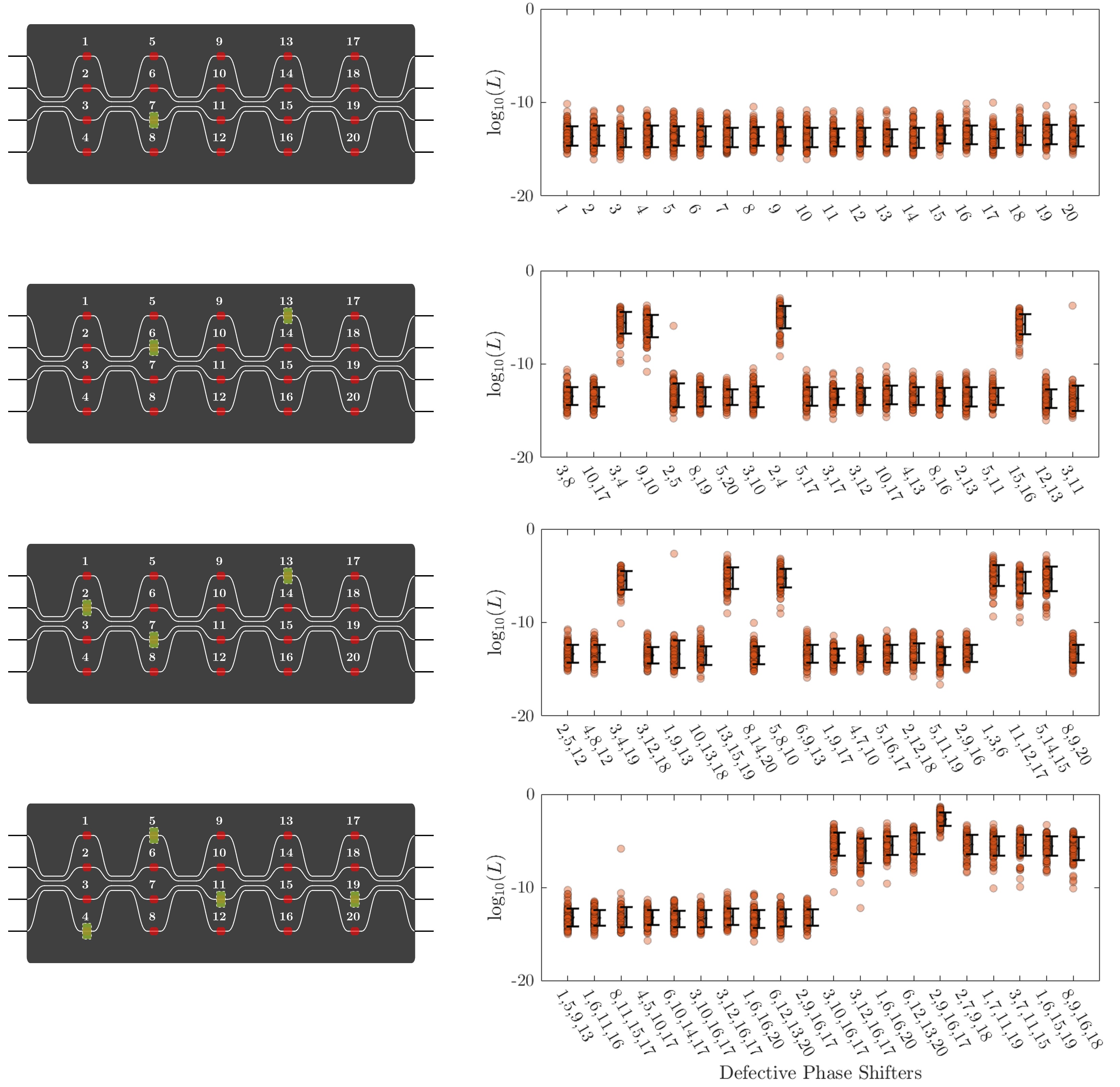}
    \caption{(Left column) Skecth of faulty phase shifters denoted in green. (Right column) The corresponding loss function for 100 randomly generated Haar target matrices for the faulty combinations in the horizontal label. A device with $N=4$ ports has been considered for $k=1$ (first row), $k=2$ (second row), $k=3$ (third row), and $k=4$ (fourth row) faulty phase shifters.}
    \label{fig_faulty}
\end{figure*}

So far, we have considered errors due to imperfections in the Jx lattice exclusively. Nevertheless, phase shifters are also susceptible to imperfections and may even be faulty in such a way that they cannot be manipulated at all. It is then required to analyze how a set of faulty phase shifters could affect the required universality of our architecture. Before doing so, it is worth remarking that we have $N(N+1)$ available phase shifters in total to reconstruct a unitary $N\times N$ unitary matrix. Thus, our setup has an over-determined number of phase shifters from the beginning, and it is plausible that even some faulty ones will not jeopardize the universality. We consider a fixed number of $k$ defective phase shifters $\{\theta_{p_1}^{(m_{1})},\ldots,\theta_{p_{k}}^{(m_{k})}\}$, where $k,p_{j}\in\{1,\ldots,N\}$. Here, $k=1$ is the case of one defective phase shifter, whereas $k=N$ is the maximum number of defective phase shifters to be considered, for that will render a device with $N^{2}$ available controllable elements. Figure~\ref{fig_faulty} shows the sketch for some faulty phase shifters and the numerical results for the loss function~\eqref{eq_L} when the remaining available phase shifters are optimized for $N=4$. In each case, we randomly select $k$ combinations of $N(N+1)=20$ phase shifters and optimize the remaining ones for $100$ randomly generated target unitary matrices. 

Figure~\ref{fig_faulty} shows the loss function for several combinations of $k$ faulty phase shifters. Notably, for $k=1$, the loss function within the tolerance error values in all the tests, regardless of the position of the faulty element. That is, the universality of the architecture is robust against a faulty element. In turn, for the cases $k=2,3,4$, numerical evidence suggests that universality is recovered whenever there is no more than one faulty phase shifter per layer. This is particularly illustrated in the last row of Fig.~\ref{fig_faulty}, where four faulty phase shifters were considered. In the initial ten testing cases, the phase shifters were distributed one per layer, while, in the last ten cases, at least two phase shifters were located in the same layer. In the former setup, the error rate is almost at the desired noise levels, except for one outlier that has an error rate of $10^{-5}$, whereas for the last ten cases, the error falls below the tolerance levels. It is important to mention that when running numerical results with 1000 targets, a few additional outliers emerge for the former setup. This could be due to the limited number of iterations used in the LMA for convergence. Increasing the number of iterations or the number of phase layers ($M>N+1$) could correct the error of these outliers, but it would significantly increase the computational time. Nonetheless, the occurrence of outliers is extremely low (almost negligible) when compared to the successful cases, and their error is still within acceptable limits, thus ensuring the reconstructed target matrices are precise enough.

\section{Conclusion}
In summary, we considered the implementation of error-correction protocols required to auto-calibrate imperfections in the construction of programmable unitary photonic circuits with interlacing architectures. The universality of the latter was already explored in previous work, where the number of phase shifter layers was proved to be $N+1$. Here, we take such an architecture further and incorporate any potential defects in the construction of the intervening photonic lattice. Although the introduced error in the DFrFT was deliberately set around one-digit percent values, the reconstructed target matrices $U_{t}$ accumulated a $10\%$ higher error as compared to the error in the DFrFT. Remarkably, the interlacing architecture allows for auto-calibration so that the so-mentioned errors can be mitigated by properly tunning the readily available phase shifters. For this task, it is required to perform a second optimization on the phase shifters by including the perturbations into the interlacing architecture. Numerical results in this regard provide evidence that the newly optimized phases bring the error back to the noise levels, revealing the desired phase corrections. This enables a more robust setup for large-scale implementations, for one can individually calibrate each of these architectures independently.

Furthermore, the architecture was shown to be resilient to errors in the event of faulty phase shifters, although it depends on the number and location of the defects. Numerical results show that the architecture is unaffected in the event of one faulty phase shifter, regardless of its position. For more than one faulty element, the universality becomes compromised whenever two faulty elements are located in the same layer. In contrast, when no more than one faulty phase shifter is present per layer, the error rate is close to the desired noise levels. This further suggests that by over-parametrizing the architecture through additional phase layers, one can allow for the presence of faulty phase shifters in the intermediate layers without jeopardizing the universality. \\

\begin{backmatter}
\bmsection{Funding}
% Note: The following will not actually appear as this is aut-generated.
This project is supported by the U.S. Air Force Office of Scientific Research (AFOSR) Young Investigator Program (YIP) Award FA9550-22-1-0189 and the City University of New York (CUNY) Junior Faculty Research Award in Science and Engineering (JFRASE) funded by the Alfred P. Sloan Foundation.

\subsection{Disclosures}
The authors declare no conflicts of interest.
%Disclosures should be listed in a separate nonnumbered section at the end of the manuscript. List the Disclosures codes identified on the \href{https://opg.optica.org/submit/review/conflicts-interest-policy.cfm}{Conflict of Interest policy page}, as shown in the examples below:
%\medskip
%\noindent ABC: 123 Corporation (I,E,P), DEF: 456 Corporation (R,S). GHI: 789 Corporation (C).
%\medskip
%\noindent If there are no disclosures, then list ``The authors declare no conflicts of interest.''

\bmsection{Data availability} Data underlying the results presented in this paper are not publicly available at this time but may be obtained from the authors upon reasonable request.

\end{backmatter}

%%%%%%%%%% If using BibTeX:
\bibliography{bibfile}

\end{document}